\documentclass{article}
\usepackage{amsmath}%
\usepackage{amsfonts}%
\usepackage{amssymb}%
\usepackage{graphicx}
\usepackage{epstopdf}
\usepackage{graphicx,graphics,color,epsfig}
\usepackage{braket}
\usepackage{cite}

\numberwithin{equation}{section}

\title { A complete and partial integrability technique of the  Lorenz system }
\author{Lazhar Bougoffa\thanks{
Al Imam Mohammad Ibn Saud Islamic University (IMSIU),
College of  Science, Department of Mathematics, P.O. Box 90950,
Riyadh 11623, Saudi Arabia. E-mail address:
lbbougoffa@imamu.edu.sa}
{}, Saud Al-Awfi\thanks{
Taibah University,
Faculty of  Science, Department of Physics, P.O. Box 30002,
Madina, Saudi Arabia E-mail address:
alawfi99@hotmail.com} 
{} and Smail Bougouffa\thanks { Physics Department, College of Science, Al Imam Mohammad ibn Saud Islamic (IMSIU) University, P.O. Box 90950, Riyadh 11623, Saudi Arabia.
E-mail address: sbougouffa@hotmail.com or sbougouffa@imamu.edu.sa}}
\begin{document}

\maketitle

\begin{abstract}
In this paper we deal with the well-known nonlinear Lorenz system that describes the deterministic chaos phenomenon. We consider an interesting problem with time-varying phenomena in quantum optics. Then we establish from the motion equations the passage to the Lorenz system. Furthermore, we show that the reduction to the third order non linear equation can be performed. Therefore, the obtained differential equation can be analytically solved in some special cases and transformed to Abel, Dufing, Painlev\'{e} and generalized Emden-Fowler equations. So, a motivating technique that permitted a complete and partial integrability of the Lorenz system is presented.
\end{abstract}
\bigskip
\noindent{\bf Keywords:} Bloch equations, Lorenz system; Abel equation; Dufing
equation; Painlev\'{e} equation; Generalized Emden-Fowler equation,
first integral
\bigskip

\section{Introduction}\label{sec1}
In the preceding papers \cite{SS07, LSS07, SS08, SB08, LS14} we investigated diverse feature of the optical Bloch equations (OBE), which are commonly employed for dynamics description in scheme consisting in electromagnetic waves, thermal bath, and the near-resonant interaction of spin or atomic transition with radiation field. Furthermore, the optical Bloch equations are explored in steady-state, in order to inspect the mean radiative forces for a two- and three-level atom. Conversely to the frequent steady state approximation, the dynamics and transient optical effects on atomic motion can be studied by complete solving of the system that governs the problem in some circumstances.

Thus, in the semi-classical approximation and rotating wave approximations~\cite{MS99, MW95}, for the two-level atom interacting with laser light, the equations of motion that are given in terms of the components of Bloch vector $(U, V, W)$, construct a system of coupled differential equations. In addition, In the case of slowly-varying amplitude $\Omega$, the equations form a non linear system. In general the solutions of this system cannot be obtained in closed forms and we have recourse to some approximate methods or numerical treatments. Thus, the equations read as

\begin{eqnarray}
\left\{
\begin{array}{rcl}
% \nonumber to remove numbering (before each equation)
   \dot{V} &=& - \kappa_2 V - \Omega(\textbf{r})W, \\
  \dot{W} &=&  \Omega({\textbf{r}})V  - \kappa_1(W +1), \\
  \dot{\Omega}&=& -\kappa_c \Omega+\omega gV,
\end{array}
\right.\label{1}
\end{eqnarray}
where for central tuning we assume that $U=0$ and $\Omega(\textbf{r})$ is the spatial dependent Rabi frequency.
The "dot" denotes the time derivative $d/dt$, $\kappa_c$ is the cavity damping rate, $\kappa_2$ is the induced dipole decay rate, $\kappa_1$ is the probability difference decay rate and $g$ is the coupling rate between the cavity field and the atom. For the standard laser threshold, in which the loss rate and the gain rate are equal, gives $\omega g=\kappa_c \kappa_2$, where $\omega$ is the field frequency. We assume that the field and the atom are in exact resonance, i.e., the detuning $(\delta= 0)$. In addition, these coupled equations are famous in literature \cite{FW04} as the Lorenz-Maxwell equations and outline a nonlinear system, which cannot always be exactly separated. Their nonlinear nature can guide to chaotic instabilities (classical chaos) in field and atomic dynamics. On the other hand, using the technique in \cite{LS14} we can pass from these equation to well known as Lorenz equations, which take the following form

\begin{eqnarray}
\left\{
\begin{array}{rcl}
\dfrac{dx}{dt}&=&a(y-x),\\
\dfrac{dy}{dt}&=&-y+cx-xz,\\
\dfrac{dz}{dt}&=&-bz+xy,
\end{array}
\right.\label{2}
\end{eqnarray}
where the constants $a>0, \ b > 0 $ and $c > 1$ are
related to the physical parameters. For example, in the previous problem of two-level atom interacting with laser field with time varying amplitude,
$a=\frac{\kappa_c}{\kappa_2}$, $c = \frac{\omega g}{\kappa_c \kappa_2},$ and $b
= \frac{\kappa_1}{\kappa_2}$.  The variables $(x, y, z)$ are expressed in terms of the normalized variables $(\Omega/\Omega_s, V/V_s, W/W_s)$ as
\begin{equation}\label{11}
    x= \beta \Omega /\Omega_{s},\quad\quad y=\beta V/V_{s}, \quad\quad z=-(W+1)/W_{s},
\end{equation}
where the index $'s'$ means the steady state solutions of equations (\ref{1}), $\beta= \sqrt{\frac{\kappa_1}{\kappa_2}(\frac{\omega g}{\kappa_c \kappa_2} -1 )}$
and time is now in units of $\frac{1}{\kappa_2}$. At this point, it can here be mentioned that  when $a > b+1$ , i.e.; $\kappa_c>\kappa_1 +\kappa_2$ , the equations (\ref{2}) can provide a complete description of disorder motion and chaos.
On the other hand, these equations are coupled ordinary differential equations and hold within two nonlinearities in the form $xz$ and $xy$.  The Lorenz equations also govern some other technical problems; i.e., in fluid mechanics dynamos, thermosyphon, electric circuits, chemical reactions and forward osmosis \cite{1, 2, 3, 4, 5}.

In general, the considered system cannot be solved analytically and we have recourse to numerical techniques, which require enough precautions on the convergence. In order to avoid these problems of convergence, we can transform these non linear coupled equations into another system in which the equations can be decoupled, and then solved independently. In this framework, we will show that the Lorenz system (\ref{2}) can be
reduced to the  Abel, Dufing, Painlev\'{e} and generalized
Emden-Fowler equations. Also, the integrable cases  and the first
integrals  of (\ref{2}) are discussed by using a reliable method,
which will be employed in a systematic fashion, for handling this
coupled nonlinear system.

\section{Reduction of  the Lorenz system to a third-order nonlinear equation  }\label{sec2}
In this section, we illustrate that the Lorenz system can be compacted to one third order differential equation. Indeed, from the first equation of (\ref{1}), we have
\begin{equation}
\label{2}y=\frac{1}{a}x'+x.
\end{equation}
Differentiating (\ref{2}) with respect to $t,$ we get
\begin{equation}
\label{3}y'=\frac{1}{a}x''+x'.
\end{equation}
Substituting  (\ref{2}) and  (\ref{3}) into the second equation of
system (\ref{1}), we obtain
\begin{equation}
\frac{1}{a}x''+x'=-\left(\frac{1}{a}x'+x\right)+cx-xz,
\end{equation}
which gives
\begin{equation}
\label{4}z=
-\frac{1}{a}\left(\frac{x''}{x}\right)-(1+\frac{1}{a})\left(\frac{x'}{x}\right)+(c-1).
\end{equation}
Hence
\begin{equation}
\label{5}z'=
-\frac{1}{a}\left(\frac{x''}{x}\right)'-(1+\frac{1}{a})\left(\frac{x'}{x}\right)'.
\end{equation}
From the third-equation of  (\ref{1}), we have
\begin{equation}
\label{6}z'+bz=xy.
\end{equation}
Substituting (\ref{2}) into the RHS of  (\ref{6}), we obtain
\begin{equation}
\label{7}z'+bz=\frac{1}{a}xx'+x^{2}.
\end{equation}
The substitution of (\ref{4}) and (\ref{5}) into (\ref{7}) leads to
the following third-order  nonlinear differential equation
\begin{eqnarray}
\label{8}&&-\frac{1}{a}\left(\frac{x''}{x}\right)'-(1+\frac{1}{a})\left(\frac{x'}{x}\right)'-\frac{b}{a}\left(\frac{x''}{x}\right)
\nonumber
\\  &&
\hspace*{10mm}-b(1+\frac{1}{a})\left(\frac{x'}{x}\right)+b(c-1)=\frac{1}{a}xx'+x^{2},
\end{eqnarray}
or
\begin{eqnarray}
\label{9}&&\left(\frac{x''}{x}\right)'+(1+a)\left(\frac{x'}{x}\right)'+b\left(\frac{x''}{x}\right)
+b(1+a)\left(\frac{x'}{x}\right)+a b(1-c)\nonumber
\\  &&
\hspace*{10mm}+xx'+ax^{2}=0.
\end{eqnarray}
We first note that Eq.(\ref{9}) has a simple solution in the form
$x=k,$ where $k$ is a constant. A simple substitution gives us
\begin{eqnarray}
k=\pm\sqrt{b(c-1)}, \ c>1,
\end{eqnarray}
which leads to the trivial solution of the system (\ref{1}) in the
form
\begin{eqnarray}
(x,\ y,\ z)=\left(\pm\sqrt{b(c-1)},\ \pm\sqrt{b(c-1)}, \ c-1\right),
\ c>1,
\end{eqnarray}
which represent the steady state solutions of the considered system. Furthermore, it is clear that three fixed points are present for $c > 1$ which corresponds to the laser threshold \cite{MW95}. As $c\rightarrow 1$, the fixed points coalesce with the origin in pitchfork bifurcation \cite{LS14}. It is clear, that the  the values of the coefficients $a, b, c$ in the Lorenz system play a crucial role in the determination of the complete integrability, partial integrability or non integrability of the system. In this way, we will discuss some important cases where the Lorenz system can be completely or partially solvable.

\section{Solutions of the Lorenz system  when $b=2a$}
Let's begin with the following case when $b=2a$. The Eq.(\ref{9}) can be written as follows
\begin{eqnarray}
\label{10}
&&\frac{d}{dt}\left[\left(\frac{x''}{x}\right)+(1+a)\left(\frac{x'}{x}\right)+\frac{1}{2}x^{2}+a(1-c)\right]
\nonumber
\\  &&
\hspace*{10mm} =-b\left[\left(\frac{x''}{x}\right)
+(1+a)\left(\frac{x'}{x}\right)+\frac{a}{b}x^{2}+a (1-c)\right].
\end{eqnarray}
If we choose $b=2a$  in the RHS of Eq.(\ref{10}), then
\begin{eqnarray}
\label{11}\left(\frac{x''}{x}\right)+(1+a)\left(\frac{x'}{x}\right)+\frac{1}{2}x^{2}+a(1-c)=C_{1}e^{-bt},
\end{eqnarray}
where $C_{1}$ is a constant of integration.
\subsection{ Case 1: $C_{1}=0$ }
In the case $C_{1}=0,$ Eq.(\ref{11}) becomes
\begin{eqnarray}
\label{12}\left(\frac{x''}{x}\right)+(1+a)\left(\frac{x'}{x}\right)+\frac{1}{2}x^{2}+a(1-c)=0.
\end{eqnarray}
The following transformation
\begin{eqnarray}
\label{13}\frac{dx}{dt}=w, \ x''=\frac{dw}{dt}=w_{x}'w
\end{eqnarray}
reduces Eq.(\ref{12}) to the  \emph{Abel differential equation of
the second kind}
\begin{eqnarray}
\label{14}w_{x}'w+(1+a)w=-\frac{1}{2}x^{3}+a(c-1)x.
\end{eqnarray}
This differential equation can be also reduced, by the introduction
of the new independent variable $\eta=-(1+a)x,$ to the \emph{Abel
differential equation of the second kind in the canonical form}
\begin{eqnarray}
\label{15}w_{\eta}'w - w= R(\eta),
\end{eqnarray}
where $R(\eta)=\frac{\frac{1}{2}x^{3}-a(c-1)x}{1+a}$ and $ \eta=-(1+a)x.$\\
Once $x$ is found then we can obtain $y$ and $z$ from the system
(\ref{1}).
%Thus, we have proved Thus
%\begin{lemma}
%Eq.(\ref{12}) can be converted to the Abel differential equation of
%the second kind in the canonical form (\ref{15}), where $w=x'.$
%\end{lemma}

\subsection{ Case 2: $C_{1}\neq0$ }
Rewrite Eq.(\ref{11}) in the following form
\begin{eqnarray}
\label{16}x''+(1+a)x'+\frac{1}{2}x^{3}+a(1-c)x-C_{1}e^{-bt}x=0.
\end{eqnarray}
Substituting the transformation
\begin{eqnarray}
\label{17}x=\lambda e^{-m t}X(\xi), \ \xi=e^{-mt},
\end{eqnarray}
where $\lambda$ and $m$  are two parameters, into Eq.(\ref{16}), we
get the following anharmonic oscillator equation
\begin{eqnarray}
\label{18}
&&X''+\left(\frac{m^{2}-m(1+a)+a(1-c)}{m^{2}}\right)\frac{1}{\xi^{2}}X+\left(\frac{3m^{2}-m(1+a)}{m^{2}}\right)\frac{1}{\xi}X'
\nonumber
\\  &&
\hspace*{20mm}+\frac{\lambda^{2}}{2m^{2}}X^{3}-\frac{C_{1}}{m^{2}}e^{-2at}\frac{1}{\xi^{2}}X=0.
\end{eqnarray}
Listed below are some special cases:
\subsubsection{ Eq.(\ref{18}) with  $a=\frac{1}{2}, \ b=1$ and $ c=0$ }
If we choose $a=\frac{1}{2}, \ b=1, \ c=0$ \cite{6,7,8} with
$m=\frac{1}{2}.$ Then Eq.(\ref{18}) reduces to \emph{the Duffing
equation} with no damping and no forcing \cite{9}
\begin{eqnarray}
X''+2\lambda^{2}X^{3}-4C_{1}X=0.
\end{eqnarray}

\subsubsection{ Eq.(\ref{18}) with $ c=1-\frac{2}{9}\frac{(1+a)^{2}}{a}$ and $a$  is an arbitrary constant }
If we choose $c=1-\frac{2}{9}\frac{(1+a)^{2}}{a},$ where $b=2a,$
with $m=\frac{1+a}{3}.$ Then Eq.(\ref{18}) reduces to \emph{the
Generalized Emden-Fowler equation} \cite{9}
\begin{eqnarray}
\label{19}X''+\frac{9\lambda^{2}}{2(1+a)^{2}}X^{3}-\frac{9\lambda^{2}C_{1}}{2(1+a)^{2}}e^{\frac{2-4a}{3}
t}X=0.
\end{eqnarray}
For example, taking as in \cite{6,7,8} $a=1,$
$\lambda=\pm\imath\sqrt{2} m$ and $C_{1}=m^{2}.$ Thus $b=2,$
$c=\frac{1}{9}$ and $m=\frac{2}{3}.$ Hence, Eq.(\ref{19}) becomes
\begin{eqnarray}
\label{20}X''(\xi)-\xi X(\xi)=X^{3}(\xi),
\end{eqnarray}
which is the\emph{ Painlev\'{e} equation}.

\begin{figure}[h]
\includegraphics[width=0.5\textwidth,height=0.5\linewidth]{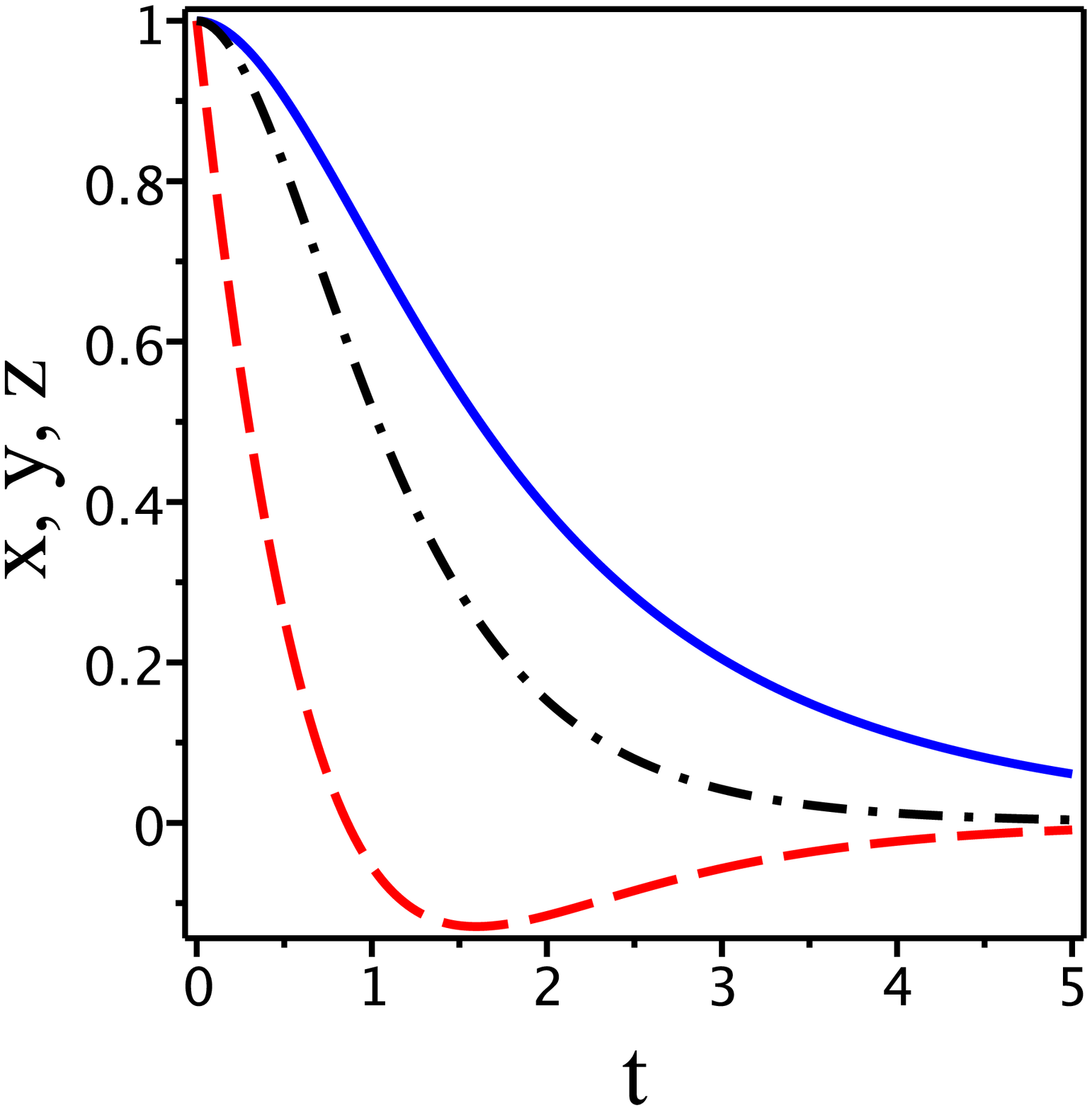}~\includegraphics[width=0.5\textwidth,height=0.5\linewidth]{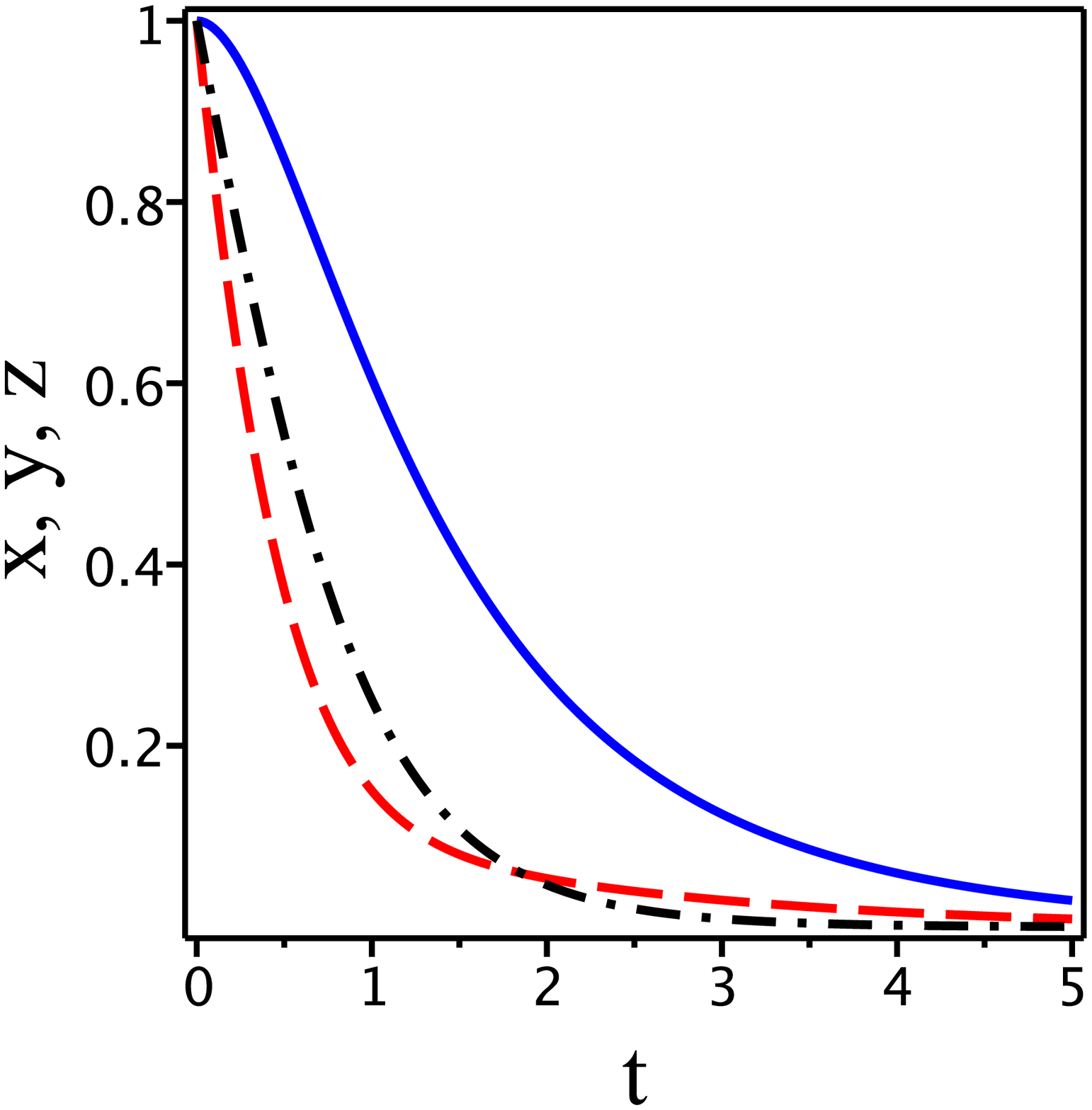}
\caption{(Color online)The solutions of the system (\ref{1}) for the initial conditions $x(0)= y(0)= z(0) = 1$, in the left box $a=0.5, b=1, c=0$ and in the right box $a=1, b=2, c=c=1-\frac{2}{9}\frac{(1+a)^{2}}{a}$, where $x(t):$ blue solid line, $y(t):$ red dashed line and $z(t):$ black dash dotted line. }
\label{Figure1}
\end{figure}

In figure (\ref{Figure1}), we represent the solutions of the system \ref{1} for the a class of initial conditions $x(0) = y(0)= z(0) =1$ and for the two previous folders for the constants $a, b, c$, where we can clearly see that the absence of the chaos, and in this cases, the Lorenz equations are complectly integrable.

\section{Solutions of the Lorenz system  when $a\rightarrow \infty$}
Now, we consider an interesting case which can be encountered in physical problems, where the constant $a$ can be very large, i.e., $a\rightarrow \infty$. From the physical problem introduced in the beginning of this paper, this situation corresponds ($a=\frac{\kappa_c}{\kappa_2}$) to the case where the critical decay rate $\kappa_c$ is too small comparatively to the induced dipole decay rate $\kappa_2$, i.e., $(\kappa_c\ll \kappa_2)$ \cite{MW95} while the other constants are arbitrary chosen. This important case is not discussed previously \cite{6,7,
8} and can present a crucial physical situations. Then  Eq.(\ref{8}) becomes
\begin{eqnarray}
-\left(\frac{x'}{x}\right)'-b\left(\frac{x'}{x}\right)+b(c-1)-x^{2}=0,
\end{eqnarray}
or
\begin{eqnarray}
\label{30}-xx''+ x'^{2}-bxx'+b(c-1)x^{2}-x^{4}=0.
\end{eqnarray}
Proceeding as before, using the  transformation (\ref{13}), we can
reduce  Eq.(\ref{30}) to \emph{the  Abel differential equation of
the second kind}
\begin{eqnarray}
\label{31}w_{x}'w=\frac{1}{x}w^{2}-b w+b(c-1)x-x^{3}.
\end{eqnarray}

\section{Integrability of the Lorenz system}

In previous sections, we have shown that the Lorenz equations can be analytically solved for some crucial classes of the parameters $a, b, c$, which correspond to an interesting physical problems. Further, It has investigated the use of the Painlev\'{e} test \cite{6, 7, 8} to find the first integrals of the Lorenz system for $c \neq 0$.  In this section, we will present an interesting technique to determine the parameter values for which the integrability and partial integrability of the Lorenz system can be performed. This technique can also be interesting for other classes of first order coupled differential equations. Indeed, we limit our investigation to the Lorenz system.\\
First, multiplying both sides of the first equation of system
(\ref{1}) by $x,$ we get
\begin{eqnarray}
\label{34}xy= \frac{1}{2a}\frac{d}{dt}(x^{2})+x^{2}.
\end{eqnarray}
From the third-equation of  (\ref{1}), we have
\begin{equation}
\label{35}xy=\frac{dz}{dt}+bz.
\end{equation}
Substituting  (\ref{35}) into (\ref{34}), we obtain
\begin{equation}
\label{36}\frac{1}{2a}\frac{d}{dt}(x^{2})+x^{2}=\lambda(t),
\end{equation}
where
\begin{equation}
\label{37} \lambda(t)= xy= \frac{dz}{dt}+bz.
\end{equation}
Solving Eq.(\ref{36}) for $x^{2},$  we get
\begin{equation}
\label{38}x^{2}= 2ae^{-2at}\int e^{2at} \lambda(t) dt+d_{1}e^{-2at},
\end{equation}
%and
%\begin{equation}
%\label{39}z= e^{-bt}\int e^{bt} \lambda(t) dt+d_{2}e^{-bt},
%\end{equation}
where $d_{1}$ is a constant of integration.\\
Thus
\begin{equation}
\label{39}x^{2}= 2ae^{-2at}\int e^{2at}\left[
\frac{dz(t)}{dt}+bz(t)\right] dt+d_{1}e^{-2at}.
\end{equation}
Hence
\begin{equation}
\label{40}x^{2}= 2a z+ \left[b(2a)-(2a)^{2}\right]e^{-2at}\int
e^{2at}z(t) dt+d_{1}e^{-2at}.
\end{equation}
Now multiply both sides of the second equation of system (\ref{1})
by $y$  and using Eq.(\ref{37}), we obtain
\begin{equation}
\label{42}\frac{1}{2}\frac{d}{dt}(y^{2})+y^{2}=(c-z)\lambda(t). %, \ G(t, z, z')=\lambda(t).
\end{equation}
Solving Eq.(\ref{42}) for $y^{2},$ we get
\begin{equation}
\label{43}y^{2}= 2e^{-2t}\int e^{2t} \left[c-z(t)\right]\lambda(t) dt+d_{3}e^{-2t},   %\phi(t, \lambda(t), d_{1} )%, \ z= \psi(t, \lambda(t), d_{2}),
\end{equation}
where $d_{3}$ is also a constant of integration.\\
Thus
\begin{equation}
\label{44}y^{2}= 2ce^{-2t}\int e^{2t}\left[
\frac{dz(t)}{dt}+bz(t)\right] dt-2e^{-2t}\int e^{2t}\left[
\frac{dz(t)}{dt}+bz(t)\right]z(t) dt+ d_{3}e^{-2t}.
\end{equation}
Hence
\begin{equation}
\label{45}y^{2}= 2cz-z^{2} +2c(b-2)e^{-2t}\int
e^{2t}z(t)dt+2(1-b)e^{-2t}\int e^{2t}z^{2}(t)dt+ d_{3}e^{-2t}.
\end{equation}
Consequently, the Lorenz system is completely integrable and the
forms
\begin{eqnarray}
\left\{
\begin{array}{rcl}
x^{2}- 2a z -d_{1}e^{-2at} =&& 2a
(b-2a)e^{-2at}\int
e^{2at}z(t) dt,\\\\
y^{2}- 2cz+z^{2}- d_{3}e^{-2t} = && 2c(b-2)e^{-2t}\\ && \times\int
e^{2t}z(t)dt+2(1-b)e^{-2t}\int e^{2t}z^{2}(t)dt.\\\\
\end{array}
\right.\label{46}
\end{eqnarray}
will lead to the first integrals of (\ref{1}).

Listed below are some special cases for different values of the parameters $a, \ b$ and $c.$
\begin{enumerate}
  \item $a=1,\ b=1$ and $c$ is an arbitrary constant \cite{6,7,8}. From
  (\ref{46}), there is a first integral in the form
  \begin{equation}
x^{2}-4cy^{2}-4z^{2}+(8c^{2}-2)z+(-d_{1}+4cd_{3})e^{-2t}=0.
\end{equation}
  \item $b=1, \ c=0$ and $a$ is an arbitrary constant \cite{6,7,8}. From
  (\ref{46}), there is a first integral in the form
  \begin{equation}
y^{2}+z^{2}- d_{3}e^{-2t}=0.
\end{equation}
  \item $b=1, \ a=\frac{1}{2}$ and $c=0$ \cite{6,7,8}. From
  (\ref{46}), the system has  two first integrals in the form
  \begin{eqnarray}
\left\{
\begin{array}{rcl}
x^{2}-  z &=&d_{1}e^{-t} ,\\\\
y^{2}+z^{2}&=& d_{3}e^{-2t}.
\end{array}
\right.
\end{eqnarray}
\end{enumerate}

\section{Conclusion}

We are concerned with the illustrious Lorenz system, which describes the deterministic chaos in different applied fields, i.e., laser phenomenon, fluid mechanics dynamos, thermosyphon, electric circuits, chemical reactions and forward osmosis, etc. The central problem that explored in this work is the passage for Lorenz system to the third order non linear differential equation. This later equation can be treated and solved for some classes of the parameter values.  In fact, we have shown that the obtained equation for some special parameter values can be reduced to the well known equations, e.g., Abel differential equation, Duffing equation, Painelev\'{e} equation, which can analytically be solved. Furthermore, a new examination, which is similar to the Painlev\'{e} analysis, is explored with the Lorenz system and it is employed to generate the previous results in an easy way. The general integrable cases of the Lorenz system and the first integrals for the Lorenz system were examined within the proposed technique.  This new approach can be extended to generate more other mathematical model that associated with chaotic phenomenon. This work is actually in progress and the results can be presented elsewhere.

\end{document}